\documentclass[a4paper]{article}

\usepackage{icrc2013}
\usepackage{amssymb}

\usepackage[english]{babel}

\title{Physics performance of the upgraded MAGIC telescopes obtained with Crab Nebula data}

\shorttitle{Performance of upgraded MAGIC telescopes}

\authors{
Julian Sitarek$^1$,
Emiliano Carmona$^2$,
Pierre Colin$^3$,
Katharina Frantzen$^4$,
Markus Gaug$^{5,6}$,
Marcos L\'opez$^7$,
Saverio Lombardi$^8$,
Abelardo Moralejo$^1$,
Konstancja Satalecka$^7$,
Valeria Scapin$^7$,
Victor Stamatescu$^1$,
Roberta Zanin$^9$,
Daniel Mazin$^3$,
Diego Tescaro$^{10,11}$
for the MAGIC Collaboration 
}

\afiliations{
$^1$ IFAE, Edifici Cn., Campus UAB, E-08193 Bellaterra, Spain \\
$^2$ Centro de Investigaciones Energ\'eticas, Medioambientales y Tecnol\'ogicas, E-28040 Madrid, Spain \\
$^3$ Max-Planck-Institut f\"ur Physik, D-80805 M\"unchen, Germany \\
$^4$ Technische Universit\"at Dortmund, D-44221 Dortmund, Germany \\
$^5$ F\'{\i}sica de les Radiacions, Departament de F\'{\i}sica, Universitat Aut\'onoma de Barcelona, E-08193 Bellaterra, Spain \\
$^6$ CERES, Universitat Aut\'onoma de Barcelona-IEEC, E-08193 Bellaterra, Spain \\
$^7$ Universidad Complutense, E-28040 Madrid, Spain \\
$^8$ INAF National Institute for Astrophysics, I-00136 Rome, Italy \\
$^9$ Universitat de Barcelona (ICC/IEEC), E-08028 Barcelona, Spain \\
$^{10}$ Inst. de Astrof\'{\i}sica de Canarias, E-38200 La Laguna, Tenerife, Spain \\
$^{11}$ Universidad de La Laguna (ULL), Dept. Astrof\'{\i}sica, E-38206 La Laguna, Tenerife, Spain
}

\email{jsitarek@ifae.es}

\abstract{
MAGIC is a system of two Imaging Atmospheric Cherenkov Telescopes located at the Canary Island of La Palma, designed to observe gamma rays with energies above ~50 GeV. Recently it has undergone an upgrade of the camera, digital trigger and readout systems. The upgrade has led to an improvement in the performance of the telescopes, especially in the lower energy range. We evaluate the performance of the upgraded MAGIC telescopes using Monte Carlo simulations and a large sample of Crab Nebula data. We study differential and integral sensitivity of the system, its angular resolution as well as its energy resolution.
}

\keywords{Cherenkov telescopes, gamma-ray astronomy}

\begin{document}
\maketitle

\section{Introduction}
MAGIC (Major Atmospheric Gamma Imaging Cherenkov) is a system of two 17 m diameter Imaging Atmospheric Cherenkov Telescopes (IACT). 
They are located at a height of 2200 m a.s.l. on the Roque de los Muchachos on the Canary Island of La Palma (28$^\circ$N, 18$^\circ$W). 
The telescopes are used for the detection of atmospheric particle showers produced by very high energy (VHE, $\gtrsim50\,$GeV) gamma rays.

In summer 2011 the readout systems of both telescopes have been upgraded and are now based on the Domino Ring Sampler (DRS) version 4 chip \cite{drs4}.
In summer 2012 the second stage of the upgrade followed with an exchange of the camera of the MAGIC~I telescope by a more finely pixelized one \cite{upgrade}. 
The new MAGIC~I camera is equipped with 1039 photomultipliers as the one used for the MAGIC~II telescope.
The upgrade of the camera has also allowed to increase the area of the trigger region in MAGIC~I by a factor of 1.7. 
We present the performance studies of the upgraded MAGIC telescopes obtained from Monte Carlo simulations and a large sample of Crab Nebula data.
The Crab Nebula \cite{whipple_crab} is considered to be ``the standard candle'' in the VHE gamma-ray astronomy as it is a bright source with so far no variability detected in the VHE energy range. 

\section{Data sample and analysis}
The used data sample consists of $22\,$h of good quality Crab Nebula data taken with the zenith angle of the source below 30 degrees.
The data were collected between October 2012 and January 2013. 
The data have been taken in wobble mode, i.e. with the source position offset by $0.4^\circ$ from the camera center.
This allows for the simultaneous background estimation from a sky position (the so-called OFF region) observed at the same distance, but on the opposite side of the camera center. 
In order to decrease the systematic uncertainties  connected with the camera inhomogeneity, the telescope pointing is modified every 20 min to swap the source and the OFF-region. 
The data have been analyzed using the standard Magic Analysis Reconstruction Software (MARS, \cite{mars}).

The analysis presented here is performed with the so-called sum image cleaning \cite{magic_pulsar, sum_cleaning}. 
Additionally, for large showers (with total charge above 750 photoelectrons) we progressively increase the cleaning thresholds. 
This procedure ensures that the image shape parameters used in the gamma/hadron separation and arrival direction estimation are computed only from the well reconstructed core of the shower. 

\begin{figure}[!t]
  \centering
  \includegraphics[width=0.49\textwidth]{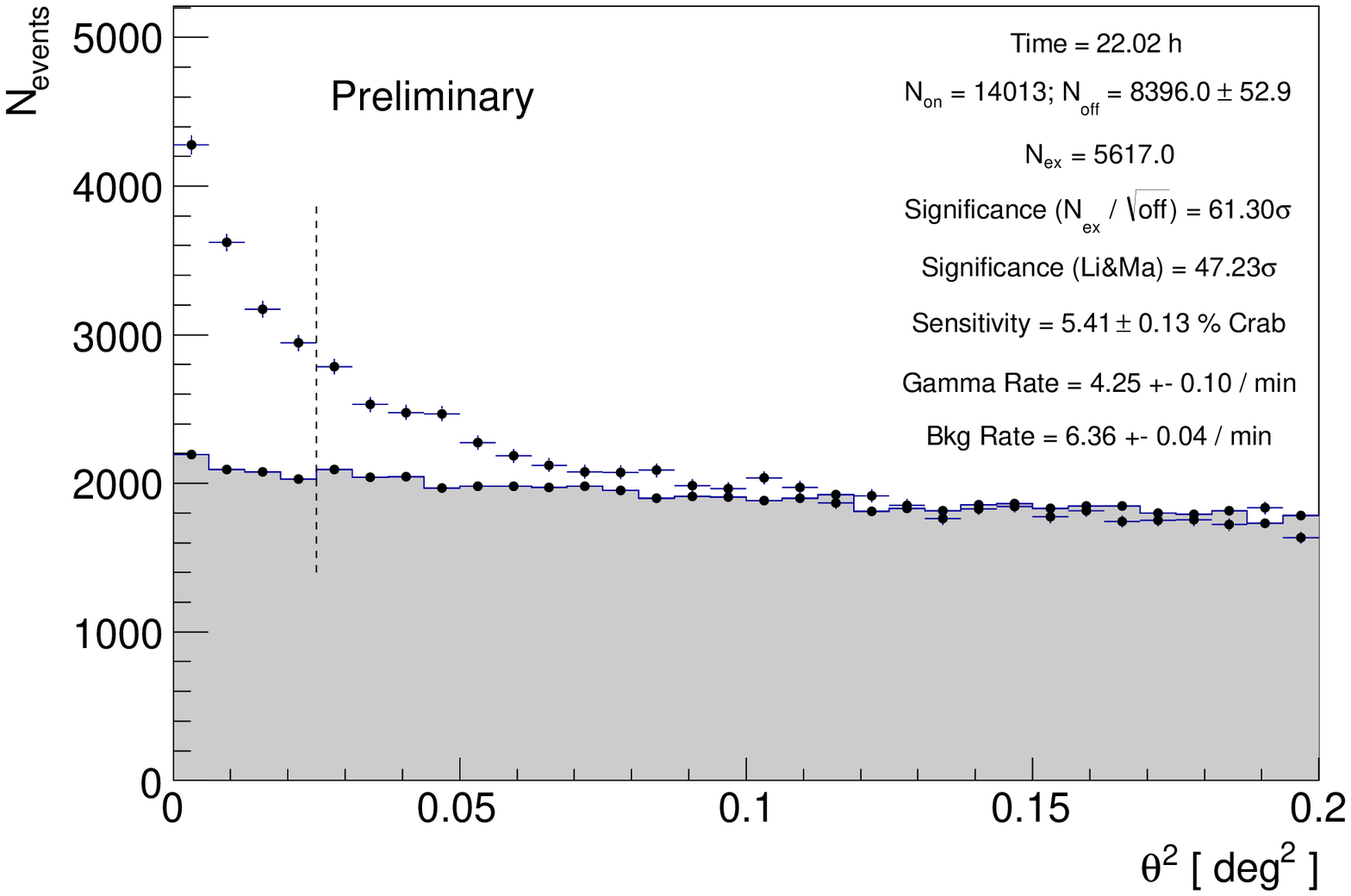}
  \includegraphics[width=0.49\textwidth]{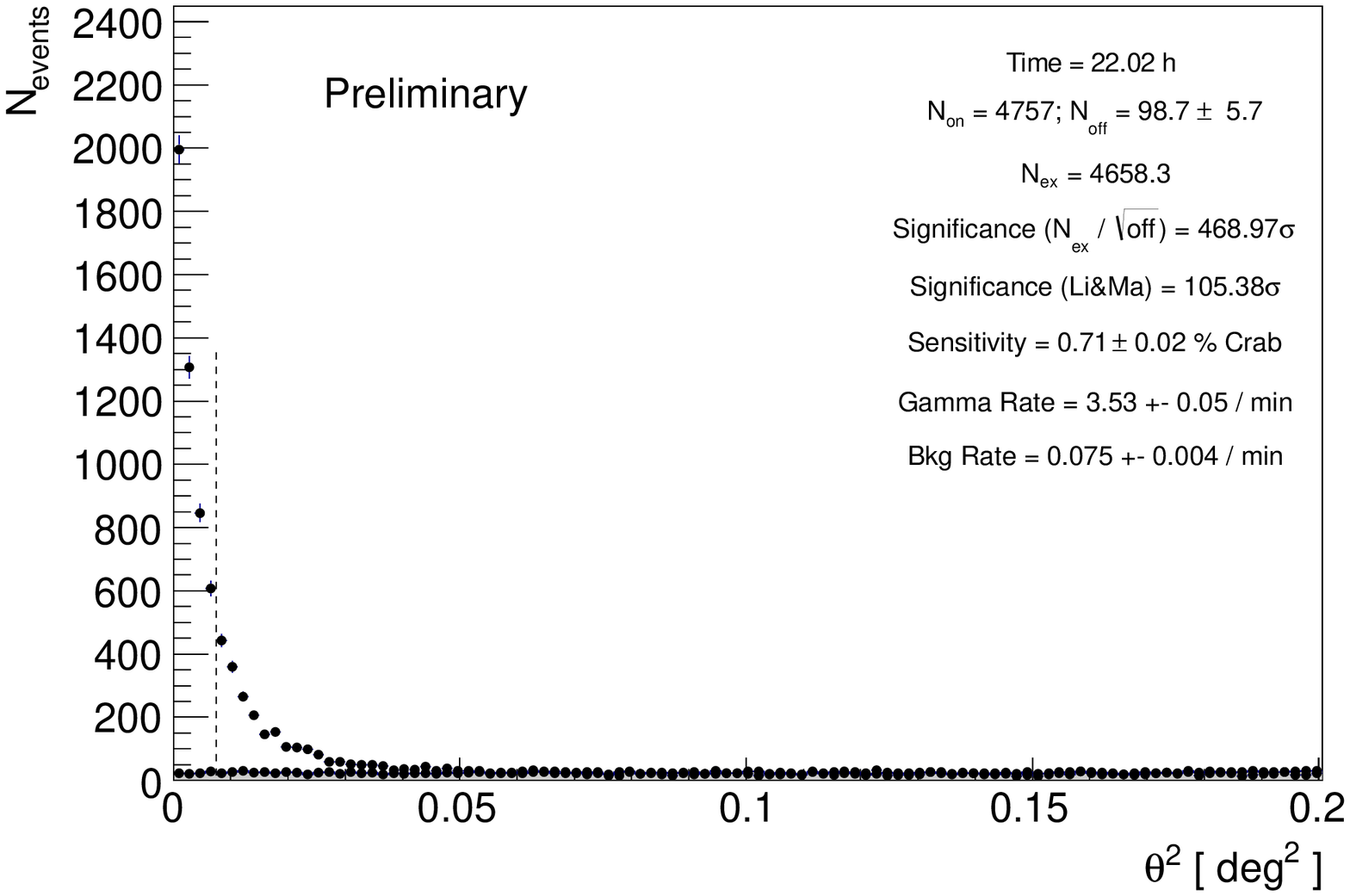}
  \caption{
Distribution of the squared angular distances between the nominal and reconstructed source position for Crab Nebula in two energy ranges: $<100\,$GeV (top panel) and $>250\,$GeV (bottom panel).
The background estimation is shown by the shaded histogram.}
  \label{fig1}
\end{figure}

In Fig.~\ref{fig1} we show the $\theta^2$ distribution, i.e. the distribution of the squared angular distances between the nominal and the reconstructed source position for the Crab Nebula.
Having a large data sample, with highly significant signal allows us to evaluate precisely the performance of the MAGIC telescopes, down to the lowest energies. 

\section{Results}

\begin{figure}[!t]
  \centering
  \includegraphics[width=0.49\textwidth]{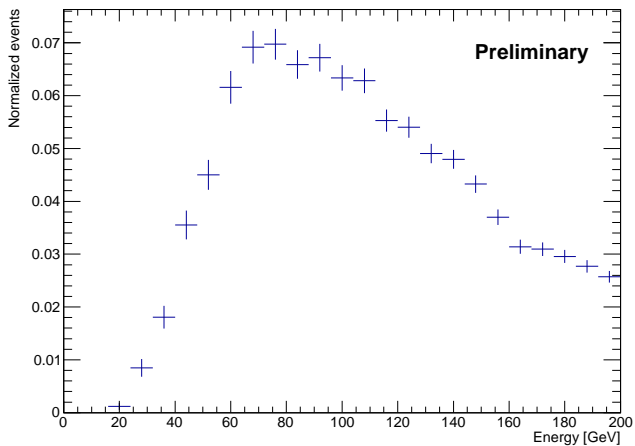}
  \caption{
Distribution of the true energy of Monte Carlo gamma-ray events after shower reconstruction, quality and gamma-ray selection cuts for a source with spectral index $-2.6$.
}
  \label{threshold}
\end{figure}

In Fig.~\ref{threshold} we show the distribution of the energy of the Monte Carlo gamma rays for the cuts used in this analysis. 
The analysis energy threshold, defined as the peak of this distribution, is $\sim 75\,$GeV.
Note however that as the peak is broad, there are many events expected with energies below this value. 
Therefore it is also possible to evaluate the performance of the telescopes not very far below the threshold. 

\subsection{Sensitivity}
In order to evaluate the sensitivity of the MAGIC telescopes after the upgrade we split the total available data sample in two subsamples of similar effective time. 
The splitting was done evenly across different observation periods and in such a way that the uniform coverage of both pointing positions was achieved in each of the two subsamples.
We use the first subsample to optimize the values of the gamma-ray selection cuts.
Afterwards we apply those cuts to the second subsample and obtain the sensitivity in an unbiased way.
We define here the sensitivity as the flux of a source which provides $N_{\rm excess}/\sqrt{N_{\rm off}} = 5$ after $50\,$h of effective observation time, where $N_{\rm excess}$ is the number of excess gamma-ray events, and $N_{\rm off}$ is the estimated background.
In order to counteract possible systematic effects on the background estimation we additionally require that $N_{\rm excess}> 5\%\,N_{\rm off}$. 
Finally, we also require $N_{\rm excess}>10$ in each energy bin. 

\begin{figure}[!t]
  \centering
  \includegraphics[width=0.49\textwidth]{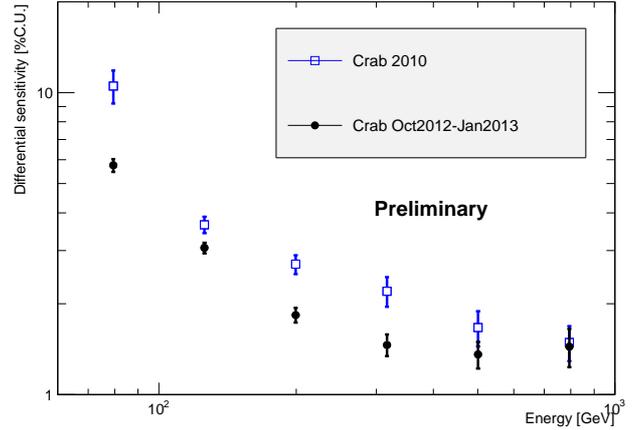}
  \caption{
Differential sensitivity in the units of percentage of the Crab Nebula flux (for $N_{\rm excess}/\sqrt{N_{\rm off}} = 5$ in $50\,$h, 5 bins per decade in energy) of the MAGIC telescopes. 
Blue open squares: the sensitivity computed with 2010 data taken before the upgrade of the readout electronics and the MAGIC~I camera \cite{perf2010}.
Black filled circles: the sensitivity of the upgraded system. 
}
  \label{fig3}
\end{figure}

Applying such conditions we investigate the differential sensitivity of the upgraded MAGIC telescopes.
We split the data into five bins per estimated energy decade and compute the sensitivity in each of them.
The differential sensitivity is shown in Fig.~\ref{fig3}.
The differential sensitivity is significantly improved, especially at the lowest energies.
For example in the energy bin $63-100\,$GeV the sensitivity improved from $(10.5\pm1.3)\%$ of the Crab Nebula flux (C.U.) to $(5.7\pm0.3)\%$ C.U.
Note that the IACT technique is affected by many systematics uncertainties, which can affect the energy scale (see \cite{perf2010} for the detailed study). 
As such a shift in the energy scale will have the largest effect around the threshold, we investigated what would be the effect on the differential sensitivity in the energy bin 63-100 GeV if the energy scale is shifted by 10\%. 
We found that even in such case, the sensitivity is $(7.4\pm0.4)\%$ C.U., which is still significantly better than the value obtained with 2010 pre-upgrade data. 
Consistent with the differential sensitivity improvement, also the integral sensitivity computed above a few hundred GeV is better, namely $\sim 0.7\%$ C.U. (see the bottom pannel of Fig.~\ref{fig1} for energy threshold of $250\,$GeV) compared with $\sim 0.8\%$ C.U. for the system prior to the upgrade \cite{perf2010}.

The sensitivities in Fig.~\ref{fig3} are given with respect to the effective observation time. 
However, one has to take into account that the 2010 data before the upgrade suffered from a $\sim10\%$ dead time in the old DRS2 based readout of the MAGIC~II telescope \cite{perf2010}.
Therefore, the sensitivity, if defined with respect to the elapsed observation time, would suffer from an additional 5\% worsening in the case of the 2010 data.
The dead time has now been reduced to a negligible amount for the upgraded readout \cite{drs4} and no longer affects the sensitivity.

\subsection{Angular resolution}
\begin{figure}[tp]
  \centering
  \includegraphics[width=0.49\textwidth]{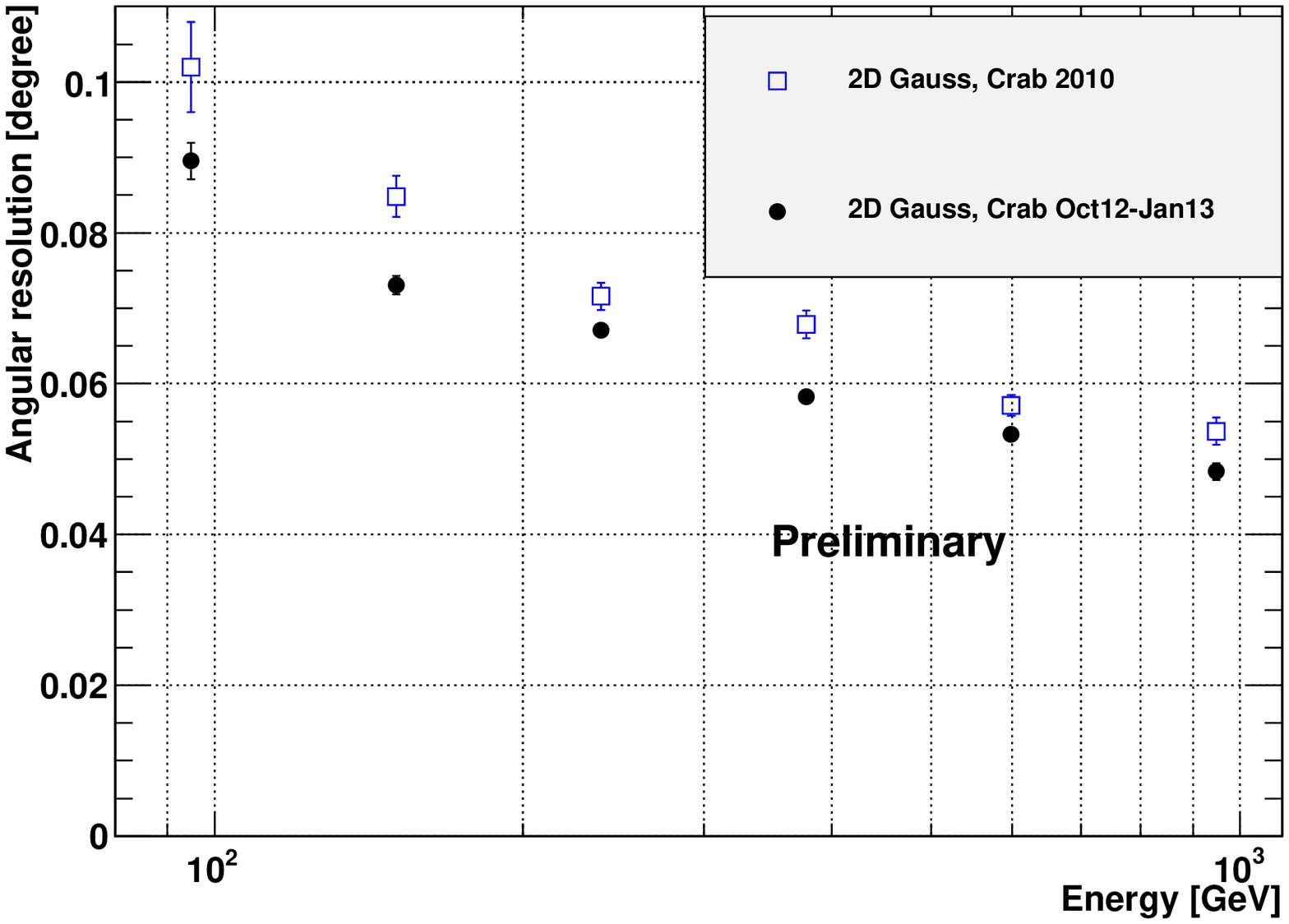}
  \includegraphics[width=0.49\textwidth]{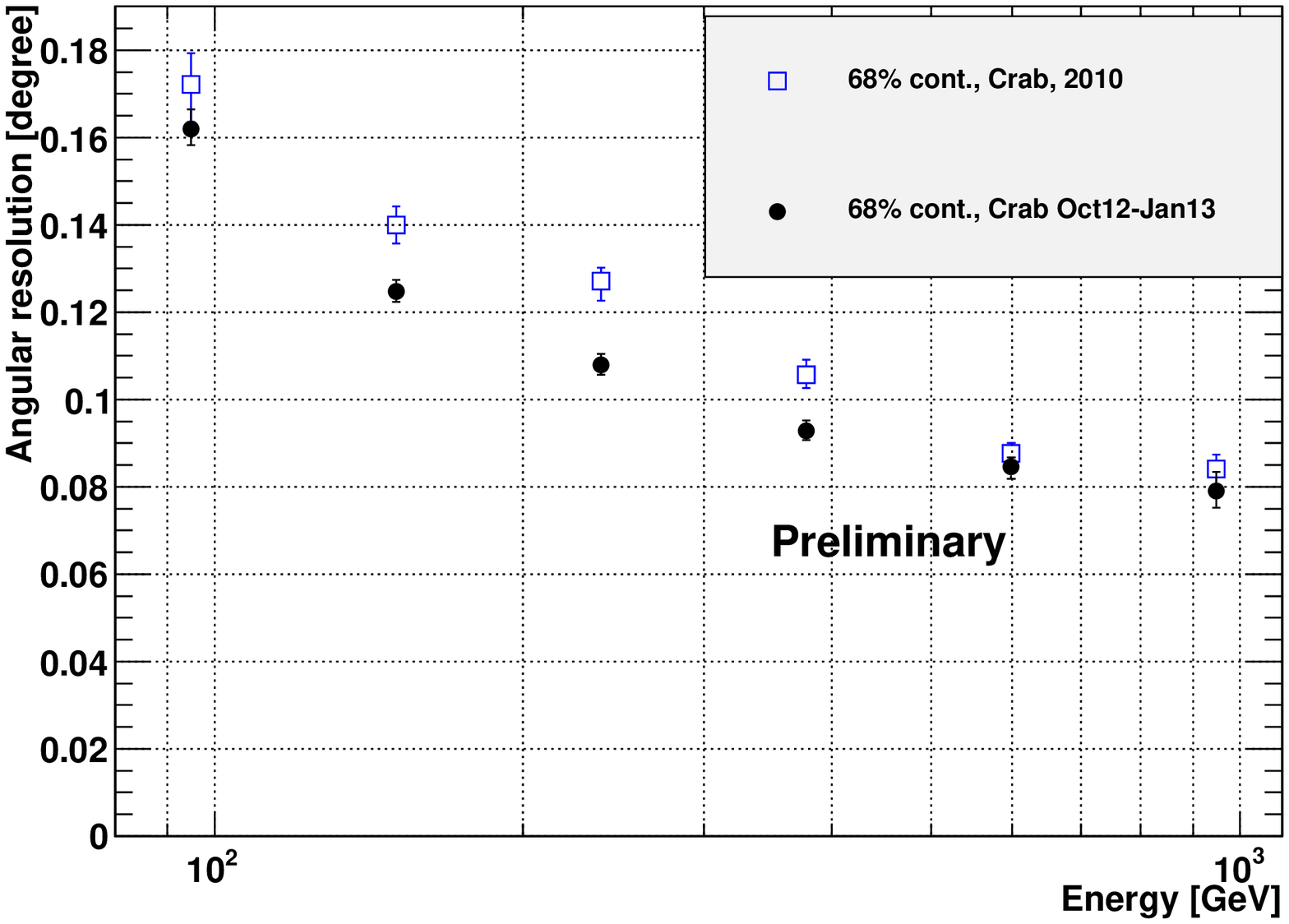}
  \caption{Angular resolution of the MAGIC telescopes as a function of energy of the gamma-ray events. 
Blue open squares: 2010 data taken before the upgrade of the readout electronics and the MAGIC~I camera \cite{perf2010}.
Black filled circles: the upgraded system. 
The top panel shows the angular resolution computed as the standard deviation of a 2-dimensional Gaussian distribution.
In the bottom panel a radius containing 68\% of the excess is shown. 
}
  \label{fig4}
\end{figure}

We investigate the angular resolution using two different methods.
In the first method we define it as the standard deviation of the 2-dimensional Gaussian fitted to the distribution of the reconstructed event directions of the gamma-ray excess events.
For a point-like source and a Gaussian point spread function response of the instrument, this corresponds to a radius containing 39\% of the gamma rays.
In the second method we directly compute the 68\% containment radius of the gamma-ray excess from the $\theta^{2}$ distribution.
In both cases we use the Crab Nebula data in different energy bins.  
The resulting angular resolution computed according to both methods is compared with the one obtained from the Crab Nebula data taken before the upgrade of the telescopes in Fig.~\ref{fig4}.
As the higher energy showers are on average better reconstructed, the angular resolution improves with energy.
After the upgrade of the MAGIC telescopes the angular resolution has improved by $\sim 5-15\%$ over the whole energy range here investigated.

\subsection{Energy resolution}
\begin{figure}[tp]
  \centering
  \includegraphics[width=0.49\textwidth]{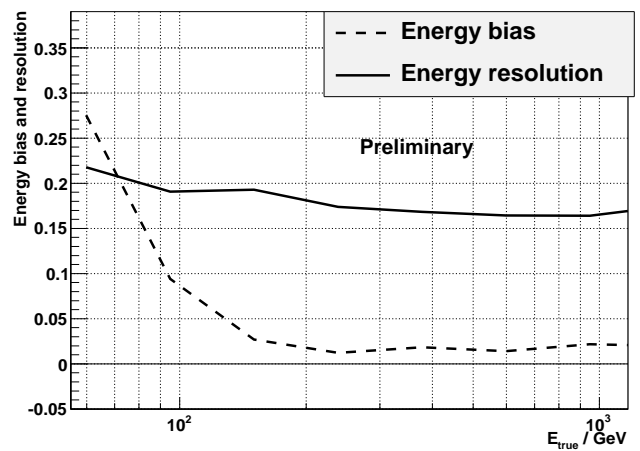}
  \caption{Energy resolution (solid line) and bias (dashed) of the MAGIC telescopes as a function of true energy of the gamma rays.}
  \label{fig5}
\end{figure}

Using Monte Carlo simulations we investigate the resolution and the bias of the energy estimation.
We bin the events in bins of true energy, and compute the following quantity $R=(E_{est}-E_{true})/E_{true}$.
For each bin of $E_{true}$ we make a distribution of $R$ and fit it with a Gaussian. 
The mean of this Gaussian is a measure of the energy bias, and its standard deviation is the energy resolution. 
Computed this way energy bias and resolution are shown in Fig.~\ref{fig5}.
The energy resolution is between 17 and 22\% in the energy range $60-1000$ GeV. 
For energies below 100 GeV there is a bias in the energy estimation due to threshold effects. 
Above 200 GeV a very small (up to a few \%) bias can appear with its magnitude and sign dependent on gamma-ray selection cuts, zenith angle range and distribution, and also the energy spectrum of the source. 
The bias is corrected in the spectral analysis of the MAGIC telescopes' data by means of unfolding \cite{mars, unfolding}.

\section{Conclusions}
The upgrade of the readout and the MAGIC~I camera has significantly improved the performance of the MAGIC telescopes, especially at the lowest energies (below 100 GeV). 
The integral sensitivity above 250 GeV is equal to $0.71\pm0.02\%$ C.U. for $50\,$h of observations. 
At those energies the angular resolution (measured as a 68\% containment radius) is $\lesssim0.1^\circ$ and the energy resolution is $\sim 18\%$.
At energies below 100 GeV the MAGIC telescopes have excellent differential sensitivity of $\sim 5.7\%$ of C.U. for $50\,$h of observations and angular resolution of $0.15^\circ$.
Such an improvement in sensitivity at the lowest energies means a reduction of the required observation time by more than a factor of 3 with respect to the performance obtained with the data taken before the upgrade. 
This makes the MAGIC telescopes an excellent instrument for the observations of VHE gamma-ray sources which can be seen only at lower energies, such as potentially Gamma-Ray Bursts, distant AGNs or pulsars. 
The performance of the telescopes at the lowest energies is expected to improve even more with the development of the new sum trigger system \cite{sum_trigger}.

\vspace*{0.5cm}
\footnotesize{{\bf Acknowledgment:}{
We would like to thank the Instituto de Astrof\'{\i}sica de
Canarias for the excellent working conditions at the
Observatorio del Roque de los Muchachos in La Palma.
The support of the German BMBF and MPG, the Italian INFN, 
the Swiss National Fund SNF, and the Spanish MICINN is 
gratefully acknowledged. This work was also supported by the CPAN CSD2007-00042 and MultiDark
CSD2009-00064 projects of the Spanish Consolider-Ingenio 2010
programme, by grant 127740 of 
the Academy of Finland,
by the DFG Cluster of Excellence ``Origin and Structure of the 
Universe'', by the DFG Collaborative Research Centers SFB823/C4 and SFB876/C3,
and by the Polish MNiSzW grant 745/N-HESS-MAGIC/2010/0.
}}

\end{document}